\documentclass[acus,keeplastbox]{jacow}
\usepackage{graphicx}  
\usepackage{subfigure} 
\usepackage{bm}        
\usepackage{amsmath}   
\usepackage{cancel} 
\usepackage{verbatim} 
\usepackage[percent]{overpic}

\usepackage[utf8]{inputenc}
\usepackage[T1]{fontenc}
\usepackage{lmodern}

\begin{document}

\title{Design Considerations for Proposed Fermilab Integrable RCS}
\author{J. Eldred and A. Valishev, FNAL, Batavia, IL 60510, USA}

\maketitle

\begin{abstract}

Integrable optics is an innovation in particle accelerator design that provides strong nonlinear focusing while avoiding parametric resonances. One promising application of integrable optics is to overcome the traditional limits on accelerator intensity imposed by betatron tune-spread and collective instabilities. The efficacy of high-intensity integrable accelerators will be undergo comprehensive testing over the next several years at the Fermilab Integrable Optics Test Accelerator (IOTA) and the University of Maryland Electron Ring (UMER). We propose an integrable Rapid-Cycling Synchrotron (iRCS) as a replacement for the Fermilab Booster to achieve multi-MW beam power for the Fermilab high-energy neutrino program. We provide a overview of the machine parameters and discuss an approach to lattice optimization. Integrable optics requires arcs with integer-pi phase advance followed by drifts with matched beta functions. We provide an example integrable lattice with features of a modern RCS - long dispersion-free drifts, low momentum compaction, superperiodicity, chromaticity correction, separate-function magnets, and bounded beta functions.

\end{abstract}

\section*{Introduction}

Integrable optics is a development in particle accelerator technology that enables strong nonlinear focusing without generating parametric resonances~\cite{Danilov}. A promising application of integrable optics is in high-intensity rings, where it is necessary to avoid resonances associated with a large betatron tune-spread while simultaneously suppressing collective instabilities with Landau damping. The efficacy of accelerator design incorporating integrable optics will undergo comprehensive experimental tests at the Fermilab Integrable Optics Test Accelerator (IOTA)~\cite{Valishev} and the University of Maryland Electron Ring (UMER)~\cite{Ruisard} over the next several years. In this paper we discuss a potential Fermilab integrable rapid-cycling synchotron (iRCS) as a high-intensity replacement for the Fermilab Booster.

At Fermilab, a core research priority is to improve the proton beam power for the flagship high-energy neutrino program~\cite{Prebys}. In the current running configuration, a 700 kW 120 Gev proton beam is delivered to a carbon-target for the NuMI beamline that supports the NOvA, MINERvA, and MINOS neutrino experiments. Next, the Proton Improvement Plan II (PIP-II) will replace the 400 MeV linac with a new 800 MeV linac that will increase the 120 GeV proton power of the Fermilab complex to 1.2 MW~\cite{PIP2}.

The next flagship neutrino experiment at Fermilab will be the LBNF/DUNE~\cite{DUNE}. The P5 Report referred to LBNF as ``the highest priority project in its lifetime'' and set a benchmark for a 3$\sigma$ measurement of the CP-violating phase over 75\% the range of its possible values~\cite{P5}. The P5 benchmark for the CP-violating phase corresponds to a 900 kt$\cdot$MW$\cdot$year neutrino exposure requirement~\cite{DUNE,Prebys}. For a 1.2 MW proton power and a 50 kt LAr detector, 15 years are required to meet that benchmark. For a 3.6 MW proton power and a 36 kt LAr detector, 7 years are required.

In order to achieve a 120 GeV proton power significantly beyond the 1.2 MW delivered by PIP-II, it will be necessary to replace the Fermilab Booster with a modern RCS~\cite{Prebys}. The Fermilab Booster is over 45 years old and faces limitations from its magnets and its RF alike. There is no beampipe inside the dipoles and the magnet laminations generate an impedance instability. The impedance instability provides $\sim$200k deceleration during transition crossing at current beam intensities~\cite{Lebedev}. The Booster dipoles are combined function magnets which constrain tunability and amplify electron cloud instabilities~\cite{Antipov}. The Booster RF cavities underwent a refurbishment process and cooling upgrade in order to achieve a 15-Hz Booster ramp rate~\cite{Pellico} but the ramp rate will not be able to exceed 20-Hz without replacing the Booster RF entirely.

\begin{figure}[htp]
\begin{centering}
\includegraphics[height=150pt, width=220pt]{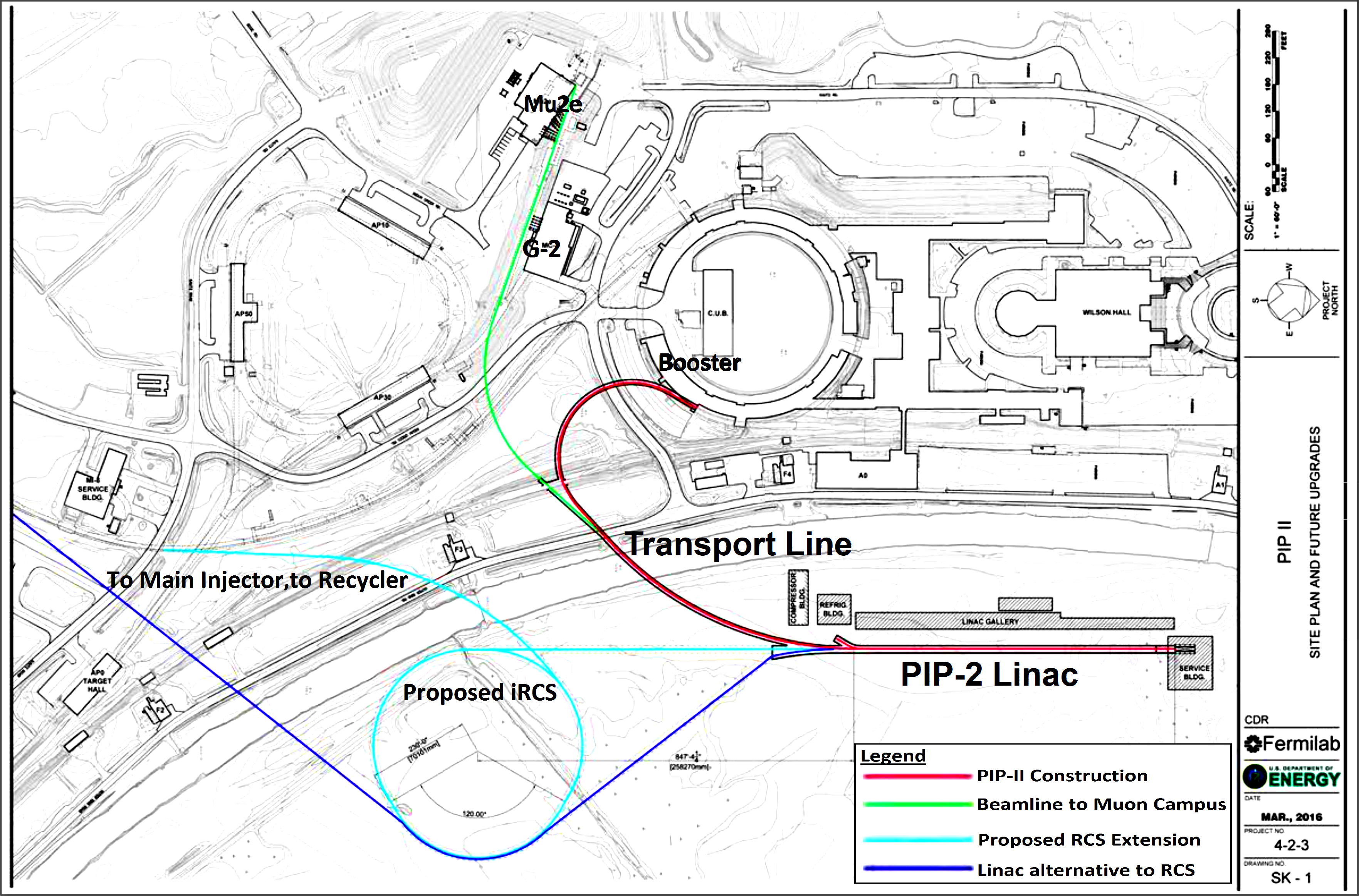}
  \caption{Site location for the proposed iRCS, relative to the PIP-II linac, muon campus, and Main Injector~\cite{Dixon}.}
  \label{Siting}
\end{centering}
\end{figure}

Figure~\ref{Siting} shows one of several siting options for an iRCS replacement for the Fermilab Booster. Neither the RCS circumference nor the injection linac length are constrained by the siting.

\section*{RCS Design Considerations}

If the PIP-II era Booster were to fill the Main Injector without slip-stacking, there would be 0.5 MW available at 120 GeV. Table~\ref{Power} shows how different parameters of a replacement RCS could modify that beam power. The bolded parameters correspond to the particular scenario that relies on integrable optics and a small increase in aperture to achieve 3.6 MW of beam power. 

\begin{table}[htp]
\centering
\caption{Multipliers on beam power relative to PIP-II Booster with boxcar stacking in the Main Injector. Bolded values shows an integrable RCS scenario that provides 3.6 MW of beam power.}
\begin{tabular}{| c | c |}
\hline
Booster-MI Beam Power & 0.5 MW \\
\hline
\hline
Laslett Tune-spread & $\Delta\nu/\Delta\nu^{(0)}$ \\
\hline
-0.11 & $\times$ 1.00 \\
-0.27 & $\times$ 2.45 \\
\pmb{-0.54} & \pmb{$\times$ 4.91} \\
\hline
Injection Energy & $\beta \gamma^{2} / (\beta_{0} \gamma_{0}^{2})$ \\
\hline
\pmb{0.8 GeV} & \pmb{$\times$ 1.00} \\
1.2 GeV & $\times$ 1.61 \\
2.0 GeV & $\times$ 3.21 \\
\hline
95\% Transverse Emittance & $\epsilon_{N} / \epsilon_{N}^{(0)}$ \\
\hline
20 $\pi$-mm-mrad & $\times$ 1.33 \\ 
\pmb{25 $\pi$-mm-mrad} & \pmb{$\times$ 1.67} \\ 
40 $\pi$-mm-mrad & $\times$ 2.67 \\ 
\hline
RCS ramp rate & $T_{MI}^{(0)} / T_{MI}$ \\
\hline
\pmb{12 Hz} & \pmb{$\times$ 0.90} \\
20 Hz & $\times$ 1.00 \\
40 Hz & $\times$ 1.09 \\
\hline
\end{tabular}
\label{Power}
\end{table}

A high-intensity RCS is constrained by betatron resonance losses incurred by a large Laslett betatron tune-spread. The Laslett tune-spread is given by
\begin{equation} \label{Las1}
\Delta \nu (z) \approx \frac{N r_{0}}{2 \pi \epsilon_{N} \beta \gamma^{2}} \left(\frac{\lambda(z)}{\langle \lambda \rangle_{z}} \right) F
\end{equation}
where $N$ is the number of particles, $r_{0}$ is the classical radius, $\epsilon_{N}$ is the normalized transverse emittance, $\lambda(z)$ is the local charge density at position $z$, and $F$ is a transverse form factor~\cite{Prebys}. Phase-space painting from PIP-II Linac will substantially improve the transverse and longitudinal beam uniformity. Eq.~\ref{Las1} can be rewritten to express the beam intensity as a function of the other three parameters:
\begin{equation} \label{Las2}
N \sim \Delta \nu_{max} \times \epsilon_{N} \times \beta \gamma^{2}
\end{equation}
where $\nu_{max}$ is the maximum Laslett tune-spread that can be sustained with minimal losses. 

If integrable optics enable a significantly higher maximum Laslett tune-spread, this would be a very cost-effective way to improve RCS performance. The subsequent section discusses how integrability impacts the RCS lattice optimization. Detailed simulations of space-charged dominated beams in integrable lattices is an ongoing work~\cite{Valishev,Bruhwiler} and the ultimate limitation on Laslett tune-spread is not fully determined.

For a fixed Laslett tune-spread, the intensity of an RCS can be improved by increasing the injection energy. Table~\ref{Power} shows how the $\beta \gamma^{2}$ parameter in Eq.~\ref{Las2} changes with an energy upgrade of the PIP-II Linac. Increasing the injection energy also reduces the transverse emittance relative to the normalized admittance, through adiabatic damping $\epsilon = \epsilon_{N}/(\beta \gamma)$.

The Fermilab Booster has a 95\% transverse admittance of 15 $\pi$-mm-mrad, but the 95\% admittance of the 5.7cm diameter beampipe is 20 $\pi$-mm-mrad without the restriction from the dipole magnets~\cite{Hassan,Prebys}. On the other hand, the 95\% transverse admittance of the Main Injector is 40 $\pi$-mm-mrad~\cite{Brown}. 

For a fixed lattice design, the transverse admittance increases quadratically with increase aperture. Either the magnet current or the accelerator circumference can be increased to compensate for the change in the magnet aperture. Table~\ref{Emit} shows the transverse emittance as a function of injection energy and aperture. Fermilab has designed RF cavities for an RCS with apertures up to 8.255cm~\cite{Hassan}.

\begin{table}[htp]
\centering
\caption{95\% Transverse admittance (in $\pi$-mm-mrad) of RCS as a function of aperture and injection energy. Asterisks indicates cases which exceed the Main Injector admittance of 40 $\pi$-mm-mrad.}
\begin{tabular}{| l | c | c | c | c |}
\hline
\multicolumn{5}{|c|}{95\% Transverse Admittance ($\pi$-mm-mrad)} \\
\hline
Injection & \multicolumn{4}{c|}{RCS Aperture} \\
Energy & 5cm & 5.7cm & 6.35cm & 8.1cm \\
\hline
0.8 GeV & 15 & 20 & \pmb{25} & 40 \\
1.2 GeV & 20 & 26 & 33 & 53* \\
2.0 GeV & 28 & 38 & 48* & 76* \\
\hline
\end{tabular}
\label{Emit}
\end{table}

Assuming conventional boxcar stacking, the impact of the RCS ramp rate on the MI beam power can be calculated by:
\begin{align}
P_{MI} &= N n_{b} E_{MI} /T_{MI} \\
T_{MI} &= T_{Ramp} + (n_{b}-1) T_{RCS}
\end{align}
where $n_{b}$ is the number of batches. As long as the Main Injector ramp remains long compared to the Main Injector fill time, the RCS ramp rate has only a modest effect on the Main Injector beam power. However, the RCS ramp rate will be an important parameter for any experiments which receive the RCS beam while the Main Injector is ramping.

An alternate RCS design with an extraction energy of 21 GeV should also be considered. Keeping the aperture and acceleration rate constant, the extraction energy of an accelerator design can be scaled by increasing the integrated dipole length, circumference, and number of RF cavities proportionately. 

For boxcar stacking, increasing the extraction energy has only a marginal effect on Main Injector beam intensity. However it should be carefully considered if a higher extraction energy would enable the RCS beam to be stacked to a greater Main Injector intensity. For example, slip-stacking is an accumulation technique currently used to double the intensity of the Main Injector, but the feasibility of slip-stacking beyond the PIP-II era is still under investigation~\cite{Brown,Eldred}. An RCS with an extraction energy of 21 GeV would avoid transition crossing in the Main Injector and that would address one of several challenges associated with high-intensity slip-stacking. An alternate accumulation approach could be transverse stacking via nonlinear-resonant injection~\cite{Giovannozzi}.

\section{IRCS EXAMPLE LATTICE}

The iRCS should incorporate the innovations in RCS design that have been developed after the Fermilab Booster~\cite{Tang}. Periodicity and bounded beta functions increase the dynamic aperture. Transition crossing can be avoided by designing the lattice with a low momentum compaction factor. Modern RCS design also uses separate-function dipole magnets and long dispersion-free drifts. In this section we show an example integrable lattice with these features.


An accelerator can achieve integrable optics with alternating sections of T-inserts and nonlinear magnets~\cite{Danilov}. The T-inserts are arc sections with $\pi$-integer betatron phase-advance in the horizontal and vertical plane. The lattice should be dispersion-free in the nonlinear section and the horizontal and vertical beta functions should be matched. A special nonlinear elliptical magnet is matched to the beta functions to provide the nonlinear focusing.

\begin{figure}[h]
\begin{centering}
  \includegraphics[height=180pt, width=220pt]{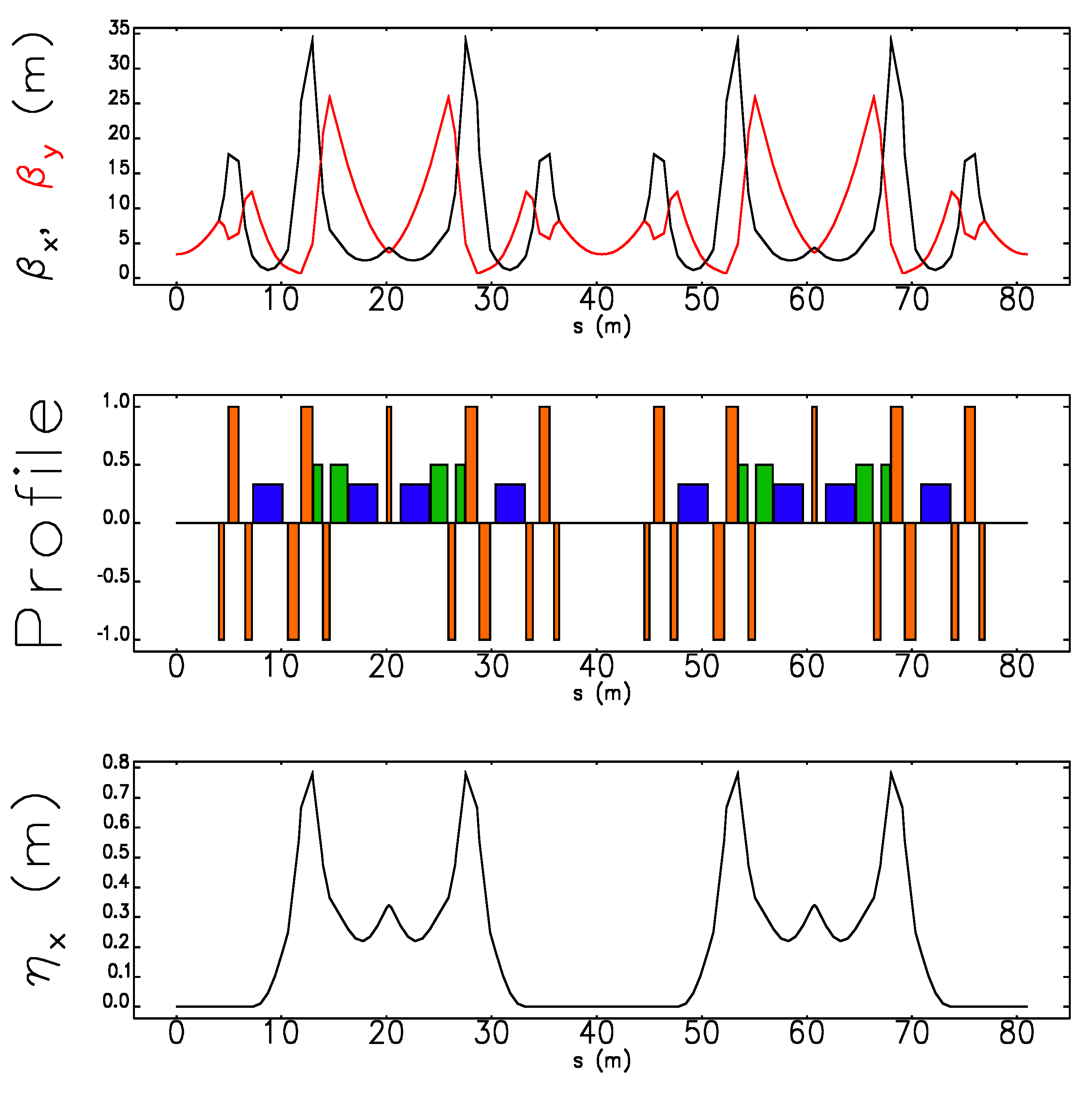}
  \caption{TWISS parameters for one of the six periodic cells. (top) Horizontal and vertical beta functions shown in black and red respectively. (middle) Location and length of magnetic lattice elements where dipoles are shown as short blue rectangles, quadrupoles shown as tall orange rectangles, and sextupoles shown as green rectangles. (bottom) Linear dispersion function.}
  \label{Lattice}
\end{centering}
\end{figure}

Figure~\ref{Lattice} shows an example iRCS lattice and Table~\ref{Param} shows the key parameters of this lattice. The lattice is composed of 12 identical achromatic arcs and dispersion-free drifts. Every other drift hosts a nonlinear insert, so the lattice forms 6 periodic cells with a T-insert section and a nonlinear insert section. The drifts in the center of each T-insert arc are used for injection, extraction, and RF acceleration.

\begin{table}[htp]
\centering
\caption{Parameters of iRCS Lattice}
\begin{tabular}{| l | c |}
\hline
Parameter & Value \\
\hline
Circumference & 486 m \\
Periodicity & 6 (12) \\
Vertical Aperture & 5 cm\\
\hline
Maximum Energy & 8 GeV \\
Bend Radius & 15.6 m \\
Peak Dipole Field & 1.25 T \\ 
Peak Quadrupole Field & 25 T/m \\ 
Peak Sextupole Field & 180 T/m$^{2}$ \\ 
\hline
Max Beta Function  & 35 m \\
Max Dispersion & 0.8 m \\
\hline
Insertion Length / Cell & 8.1 m \\
Total Insertion Length & 97 m \\
Single Dipole Length & 2.8 m \\
Number of Dipoles & 48 \\
Number of Quadrupoles & 156 \\ 
Number of Sextupoles & 48 \\ 
\hline
Momentum Compaction & 2$ \times 10^{-3}$ \\
Extraction Phase-Slip Factor & -6$ \times 10^{-3}$ \\
Betatron Tunes & 19.7 \\
Linear Chromaticities & -10 \\
Second-order Chromaticities & 50 \\
\hline
\end{tabular}
\label{Param}
\end{table}

This example lattice is compatible with the 8-GeV lattice described in the previous section. By increasing the circumference and number of cells, the 8-GeV lattice can easily be scaled up to 21-GeV lattice with a lower momentum compaction factor.


The effect of linear chromaticity on integrable motion was examined in \cite{Webb} and the effect of nonlinear chromaticity was examined in \cite{Cook}. Chromaticity is compatible with integrability if the horizontal and vertical chromaticities are matched. In the example iRCS lattice the chromaticities were matched and reduced in both the first and second order. The chromaticity values shown in Fig.~\ref{Param} are the result of correction by the sextupole magnets (green in Fig.~\ref{Lattice}).

To preserve integrability, sextupole magnets should also be located so that their effect cancels harmonically (separated by a $\pi$-odd phase-advance)~\cite{Webb}. To maintain the flexibility of the early design, this constraint was not imposed on the example lattice shown here. This constraint can be met by requiring a $\pi$-odd phase-advance for the 12 linear-periodic cells or by combining into 6 complex linear-periodic cells.

\section{SUMMARY \& FUTURE WORK}

To achieve multi-MW beam power for the Fermilab high-energy program, an integrable RCS replacement for the Fermilab Booster is an option that merits careful scrutiny. In this paper we explore some of the preliminary design concerns and provide an example integrable RCS lattice design. Experimental and numerical work on the interaction between integrable optics and space-charge dominated beams is still ongoing~\cite{Valishev,Ruisard,Bruhwiler}.

Upcoming efforts will study the space-charge dynamics of the lattice, integrate sextupoles with harmonic cancellation, and investigate longitudinal and transverse stacking schemes. The design of the high-power H$^{-}$ stripping foil and injection chicane will also be developed.

\end{document}